\documentclass[useAMS,usenatbib]{mn2e}

\usepackage{graphicx}
\usepackage{amsmath}

\usepackage{color}

\title[Gamma-rays from nebulae around binary systems containing pulsars]
{Gamma-rays from nebulae around binary systems containing energetic rotation powered pulsars}
\author[W. Bednarek \& J. Sitarek]
{W. Bednarek$^1$ \& J. Sitarek$^2$\\
$^1$Department of Astrophysics, The University of \L \'od\'z,
ul. Pomorska 149/153, 90-236 \L \'od\'z, Poland,
bednar@astro.phys.uni.lodz.pl\\
$^2$IFAE, Edifici Cn., Campus UAB, E-08193 Bellaterra, Spain, jsitarek@ifae.es}
\begin{document}

\date{Accepted . Received ; in original form }

\pagerange{\pageref{firstpage}--\pageref{lastpage}} \pubyear{2007}

\maketitle

\label{firstpage}

\begin{abstract}
We consider nebulae which are created around binary systems containing rotation powered pulsars and companion stars with strong stellar winds. It is proposed that the stellar and pulsar winds have to mix at some distance from the binary system, defined by the orbital period of the companion stars and the velocity of the stellar wind. The mixed pulsar-stellar wind expands with a specific velocity determined by the pulsar power and the mass loss rate of the companion star.  Relativistic particles, either from the inner pulsar magnetosphere and/or accelerated at the shocks between stellar and pulsar winds, are expected to be captured and isotropized in the  reference frame of the mixed wind. Therefore, they can efficiently comptonize stellar radiation
producing GeV-TeV $\gamma$-rays in the inverse Compton process. We calculate the $\gamma$-ray spectra expected in such scenario for the two example binary systems: J1816+4510 which is the redback type millisecond binary and LS~5039 which is supposed to contain energetic pulsar. It is concluded that the steady TeV $\gamma$-ray emission from J1816+4510 should be on the 100 hr sensitivity limit of the planned Cherenkov Telescope Array, provided that $\epsilon\sim 10\%$ of the rotational energy lost by the pulsar is transferred to TeV electrons. On the other hand, the comparison of the predicted steady TeV $\gamma$-ray emission, expected from $\gamma$-ray binary LS~5039, with the observations of the TeV emission in a low state, reported by the H.E.S.S. Collaboration, allows us to put stringent upper limit on the 
product of the part of the hemisphere in which the mixed pulsar-stellar wind is confined, $\Delta_{\rm mix}$, and the energy conversion efficiency, $\epsilon$, from the supposed pulsar to the TeV electrons injected in this system, $\Delta_{\rm mix}\cdot \epsilon <1\%$. This lower limit can be understood provided that either the acceleration efficiency of electrons is rather low ($\epsilon\sim 1\%$), or the parameters of the stellar wind from the companion star are less extreme than expected, or the injection/acceleration process of electrons occurs highly anisotropically with the predominance towards the companion star.

\end{abstract}

\section{Introduction}

Some binary systems contain rotation powered pulsars which are energetic enough to prevent accretion of matter from stellar winds of companion stars. These pulsars can create strong pulsar winds which collide with the winds of the companion stars. As a result, a shock structure appears within the binary system on which particles can be accelerated to relativistic energies. A few types of such binaries have been discovered up to now. The most famous is a class of "Black Widow" binaries in which energy output from the millisecond pulsar 
can even partially evaporate the companion star. The companion stars in these systems are characterised by small masses, below $\sim 0.1$M$_\odot$, and relatively strong winds induced by energy released by the pulsar. In another type of such binaries, the so-called "Redback" binary systems, the energetic millisecond pulsar forms a binary system with a low mass $\sim 0.2$ M$_\odot$ star (see \citealp{ro11}). Up to now, only one binary system has been discovered in the Galaxy to contain a classical $\gamma$-ray pulsar on a close orbit around the massive, Be type, star (i.e. PSR 1259-63/SS 2883, \citealp{jo92,jo94}). It has been proposed that other binary systems can also contain ejecting classical pulsars, e.g. LSI 303 +61 \citep{mt81}, Cyg X-3 \citep{hg90} or LS~5039 \citep{dub06}. These binaries emit $\gamma$-rays in the GeV-TeV energy range, clearly modulated with the period of the binary system \citep{al09,ah06}. In the most popular scenario, these $\gamma$-rays are produced by relativistic electrons which up-scatter stellar radiation in the Inverse Compton process,  \citep[e.g.][]{mt81,ta97,bed97}. Since the stellar radiation field is anisotropic for electrons ejected from the pulsar or accelerated at the shock, the emission has to be modulated with the period of the binary system. However, a part of relativistic electrons in the pulsar wind can escape from the binary system without significant interactions, i.e. those moving close to the outward direction. In the case of low mass millisecond pulsar binaries, most of the accelerated electrons can escape radially from the vicinity of the companion star, due to relatively weak radiation field.
Farther from the binary systems, the winds from the pulsar and the companion star have to mix due to the rotation of the binary system.
As a result, it is expected that such mixed pulsar-stellar winds move together with some specific velocity determined by the energy output from the pulsar and the material content of the stellar wind. 
Therefore, nebulae around binary systems containing rotation powered pulsars should have different structures than nebulae around isolated pulsars. 
We propose that such binary systems should be surrounded by nebulae composed of the mixture of the pulsar and stellar winds which expands with non-relativistic velocity. Relativistic particles are expected to be captured and isotropized in the reference frame of this wind.

In this paper we consider radiation processes occurring in such nebulae. We calculate the $\gamma$-ray spectra produced by relativistic electrons in the Inverse Compton (IC) scattering of thermal radiation from the companion star. In contrast to the modulated $\gamma$-ray emission from the interior of the binary system, we predict the presence of a steady component of the $\gamma$-ray emission produced in the nebula around the binary system.  Discovery of such additional component by the future Cherenkov telescopes (e.g. Cherenkov Telescope Array, CTA, \citealp{ac11}) from the binaries 
containing pulsars should allow to derive important parameters determining the acceleration processes in such binaries and properties of the emission region. The comparison of the predicted high energy emission with the available  observations of the TeV $\gamma$-ray emission in the low state from already known $\gamma$-ray binaries should allow to provide additional constraints on the acceleration efficiency of particles occurring around the rotation powered pulsars in the binary systems.

\section{Nebulae around binary systems}

\begin{figure*}
\vskip 7.5truecm
\includegraphics{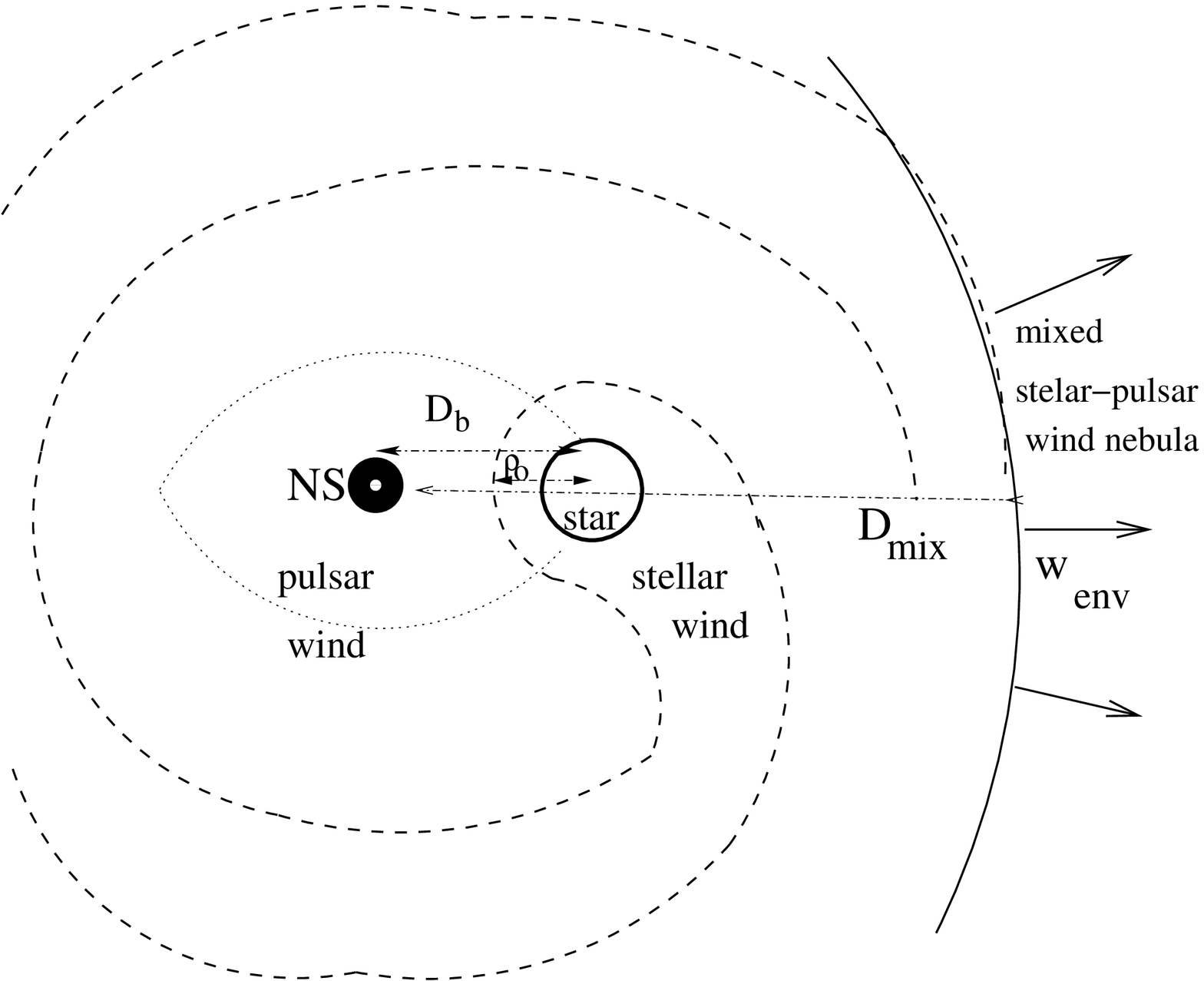}
\includegraphics{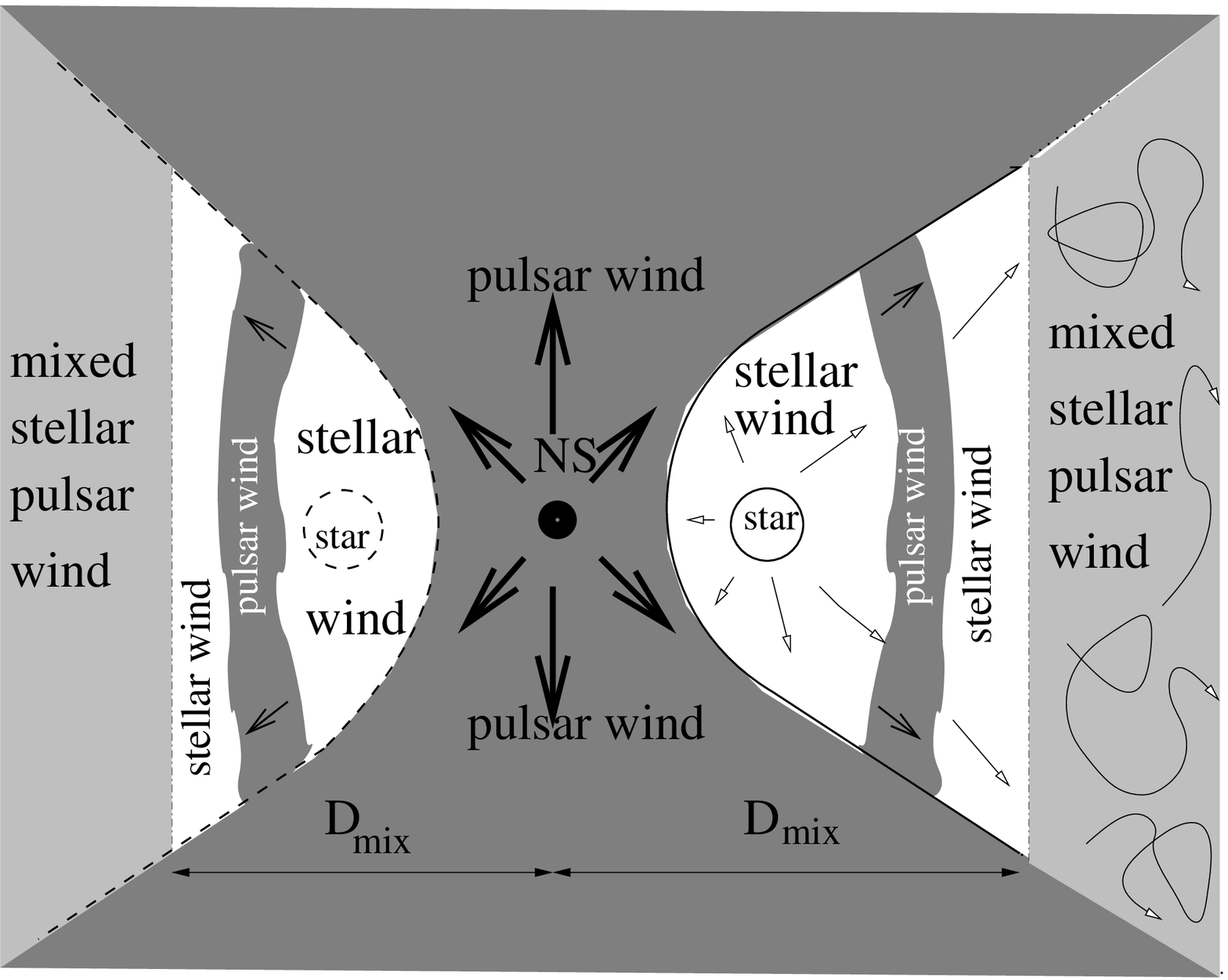}
\caption{Schematic representation (not to scale) of the vicinity of the binary system containing an energetic pulsar and a companion star (seen from the reference frame of the pulsar). The top view (i.e. from the pole of the binary system) is shown on the left and the back view 
(from the side of the binary plane) is shown on the right. 
On the left: The dotted line shows the orbit of the star.
The pulsar creates energetic wind which collides with the stellar wind.
As a result, a double shock structure is created (marked by the dashed curve). The stellar (pulsar) wind creates a cometary tail which interacts with the pulsar (stellar) wind. At some distance from the binary system (signed by $D_{\rm mix}$) both winds mix together. The mixed stellar-pulsar wind propagates with the common velocity estimated from Eq.~\ref{eq3}. Relativistic particles accelerated by the pulsar move rectilinear in the pulsar wind up to the shock structure. In this region many of them are not able to interact with the stellar radiation due to the small interaction angles. However, in the mixed stellar-pulsar wind they are isotropized. They start to lose efficiently energy on comptonization of the stellar radiation. Moreover some particles can be also accelerated at the shock between the pulsar and stellar winds in the Fermi acceleration process. On the right: This same picture but viewed from the plane of the binary system for the case of the parameter $\eta>1$. In this case the pulsar wind can escape freely only along the polar regions of the binary system. The stellar and pulsar winds have to mix at the distance $D_{\rm mix}$ in the region of the plane of the binary system.
}
\label{fig1}
\end{figure*}

As we have outlined above we consider specific type of nebulae around binary systems containing rotation powered pulsars and companion stars with 
substantial mass loss rate. In fact, in such type of binaries
a double shock structure appears already within the binary system in collision of the pulsar and stellar winds. Its parameters are determined by
the energy loss rate of the pulsar and the power of the stellar wind.
In the case of spherical winds, the distance of the shock from the star is defined by the parameter, 
$\eta = (L_{\rm pul}/c)/{\dot M}_{\rm w} v_{\rm w}\approx 0.05 L_{36}/(M_{-7}v_8)$ \citep{gw87}, where
the pulsar wind power is $L_{\rm pul} = 10^{36}L_{36}$ erg/s, the stellar wind mass loss rate is ${\dot M}_{\rm w} = 10^{-7}M_{-7}$ M$_\odot$/yr and its velocity is $v_{\rm w} = 10^8v_8$ cm/s.
The closest, radial distance of the shock from the pulsar is then given by $\rho_{\rm o} = D_{\rm b}\sqrt{\eta}/(1 + \sqrt{\eta}$), where $D_{\rm b}$ is the separation of the binary stars.
For the example scaling parameters applied above, the shock should be bound around the pulsar. In the case of binary systems containing energetic millisecond pulsars, the shocks are usually bound around the low mass companion stars due to relatively small power of the stellar winds in respect to power of the winds around massive stars. 
These shock structures can be much more complicated in the case of non-spherical stellar winds of massive stars of the Be type (see \citealp{sb08}). In this last case the location of the shock structure can change drastically depending on the location of the pulsar within the polar or equatorial stellar wind.

The shock structure tends asymptotically to a cone characterised by a half-opening angle $\psi = 2.1 (1 - \eta_{\rm m}^{2/5}/4)\eta_{\rm m}^{1/3}$ rad, where $\eta_{\rm m} =$ min($\eta,\eta^{-1}$) (see \citealp{bd01,eu93}). However, due to the rotation of the binary system,
the cone should be bound at large distances from the stars (e.g. Dubus 2006, Bosch-Ramon et al.~2012) provided that the velocity of the stellar wind is large in respect to the rotational velocity of the companion star.
For $\eta > 1$ the pulsar wind completely overtakes the stellar wind, creating a torus like structure in the equatorial plane of the binary system (see. Fig.~1 for schematic geometry).
In this case, the pulsar wind can only propagate freely along the  polar regions of the binary system. 
In the opposite case $\eta < 1$, the stellar wind overtakes the pulsar wind. However, again in the equatorial region the pulsar wind mixes with the stellar wind as in the previous case.
There are significant differences between the mixed pulsar-stellar wind in the equatorial region and the freely expanding pulsar wind in the polar region. Relativistic electrons are isotropized in the reference frame of a relatively slowly expanding mixed wind.
These electrons can interact efficiently with the thermal radiation from the nearby companion star. Relativistic electrons in the pulsar wind,
expanding in polar directions, move rectilinear from the binary system.
These electrons can be isotropized only far away from the binary 
system where the density of stellar photons is already very low (as in the case of PWNe around isolated classical pulsars).
These electrons are not expected to produce efficiently $\gamma$-rays
by scattering stellar radiation.
They finally scatter the Microwave Background Radiation and the infrared radiation in the Galaxy producing extended $\gamma$-ray sources which are more difficult to detect with the Cherenkov telescopes.
If the parameter $\eta$ is lower than unity, then the equatorial regions of the binary system are dominated by the mixed pulsar-stellar wind but the 
polar regions are dominated by the stellar wind. In this case all electrons in the pulsar wind are expected to be isotropized at some distance from the binary system.

We calculate the factor which determines a part of the hemisphere (the solid angle divided by $4\pi$) in which the mixed pulsar-stellar wind (the toroidal structure in the plane of the binary system, see Fig.~1) is confined. This factor is calculated for the known value of the half opening angle, $\psi$, of the shock created within the binary system in collisions of these two winds. 
This part of the hemisphere is equal to $\Delta_{\rm mix} = \Delta\Omega_{\rm mix}/4\pi = (1 - \cos\psi)$, where $\Delta\Omega_{\rm mix}$ is the solid angle in which the mixed pulsar-stellar wind is confined. Both winds are expected to mix completely at the distance from the binary system which depend on the velocity of the stellar wind and the rotational period of the binary system,
\begin{eqnarray}
D_{\rm mix}\approx \tau_{\rm orb}v_{\rm w}\approx 8.6\times 10^{12}\tau_{\rm d}v_8~~~{\rm cm},
\label{eq1}
\end{eqnarray}
\noindent
where the period of the binary system is $\tau_{\rm orb} =1\tau_{\rm d}$ days. In the above formula we assumed that the stellar wind velocity is much greater than orbital velocity of the star which is usually the case for the considered binaries. As a result of the mixing process, the pulsar wind becomes uploaded with the stellar wind matter.
The mixture of both winds starts to move together with the common velocity, $w_{\rm env}$, estimated from the energy conservation,
\begin{eqnarray}
L_{\rm pul} + L_{\rm sw} = {\dot M}w_{\rm env}^2/2,
\label{eq2}
\end{eqnarray} 
\noindent
where $L_{\rm sw} = {\dot M}_{\rm w}v_{\rm w}^2/2$ is the kinetic power of the stellar wind.
Here we neglect the energy loss of the bulk motion of the mixed winds on the isotropization of the direction of the particles in it.
If the stellar wind power dominates over the pulsar wind power, then $w_{\rm env}\approx v_{\rm w}$. In the opposite case, 
\begin{eqnarray}
w_{env} = \sqrt{2L_{\rm pul}/{\dot M}}\approx   5.4\times 10^8 (L_{36}/M_{-7})^{1/2}~~~{\rm cm/s},
\label{eq3}
\end{eqnarray}
\noindent
resulting in a non-relativistic bulk motion of the mixed winds for the assumed above example parameters.

Relativistic electrons accelerated in the inner binary system, either by the pulsar itself or on the shock formed in the pulsar and stellar winds collision, are immersed in the mixed stellar-pulsar wind. We estimate the maximum energies of the electrons for which they are captured and confined by the mixed wind magnetic field. It is assumed that the magnetic field in the mixed wind
is determined by the magnetic field in the stellar wind. It has a dipole structure only very close to the stellar surface. Farther out, it has a radial structure and finally reaches a toroidal structure due to the rotation of the star (for details see \citealp{um92}). The magnetic field strength with the toroidal structure can be approximated as a function of the distance from the star by the following formula,
\begin{eqnarray}
B(D)\approx 0.01B_{2}R_{11}/D_{13}~~~{\rm G},
\label{eq4}
\end{eqnarray}
\noindent
assuming that the transition between the region of the toroidal and the radial magnetic field occurs at the distance of $\sim 10$ stellar radii (where the radius of the star is $R_\star = 10^{11}R_{11}$ cm). In this formula, the surface magnetic field of the star is $B_\star = 100B_{2}$ G, and the distance from the star is expressed as $D = 10^{13}D_{13}$ cm.

We assume that relativistic electrons are captured by the wind (and advected with the wind) if their Larmor radius, $R_{\rm L}$, is lower than the characteristic distance scale given by the distance from the massive star, i.e. $R_{\rm L}\approx 3\times 10^{9}E_{\rm TeV}/B(D_{\rm mix}) \mathrm{cm} < D_{\rm mix}$, where $E = 1E_{\rm TeV}$ TeV is the energy of an electron, and $B(D_{\rm mix})$ (measured in Gauss) is the local magnetic field strength in the mixed wind. Based on this condition, we conclude that electrons with energies, 
\begin{eqnarray}
E < 30B_{\rm 2}~~~{\rm TeV},
\label{eq5}
\end{eqnarray}
\noindent
should be captured in the mixed stellar-pulsar wind.
 
Electrons, captured in the magnetic field of the mixed stellar-pulsar wind, are isotropized in the reference frame of the relatively slow wind.
They lose energy on the synchrotron radiation and on the IC scattering of the stellar radiation. 
The importance of the synchrotron energy losses can be evaluated by comparing the synchrotron cooling time scale with the advection time scale with the velocity of the winds. The advection time scale can be estimated from, 
\begin{eqnarray}
\tau_{\rm adv} = D/w_{\rm env} = 10^5D_{13}/w_8~~~{\rm s},
\label{eq6}
\end{eqnarray}
\noindent
where $v_{\rm env} = 10^8w_8$ cm/s, and
the synchrotron time scale from, 
\begin{eqnarray}
\tau_{\rm syn} = E_{\rm e}/{\dot E}_{\rm syn}\approx 290/(B^2E)~~~{\rm s},
\label{eq7}
\end{eqnarray}
\noindent
where ${\dot E}_{\rm syn} = (4/3)cU_{\rm B}\sigma_{\rm T}E_{\rm e}^2/m_{\rm e}^2\approx 3.5\times 10^{-3}B^2E^2$ TeV/s, $m_e=511\,$keV and electron energy $E$ is expressed in TeV. By comparing Eq.~\ref{eq6} and~\ref{eq7}, we get the limit on the electron energy above which the synchrotron energy loss time scale is shorter than the advection time scale,
\begin{eqnarray}
E_{\rm adv}^{\rm syn}\approx 29w_8D_{13}/(B_2R_{11})^2~~~{\rm TeV}.
\label{eq8}
\end{eqnarray}
\noindent
This critical electron energy is equal to $E_{\rm e}\approx 130\tau_{\rm d}v_8(L_{36}/M_{-7})^{1/2}/(B_2R_{11})^2$ TeV, for the mixture distance of the winds given by Eq.~\ref{eq1} and the velocity of the mixed wind given by Eq.~\ref{eq3}.
We conclude that electrons with TeV energies are not able to lose efficiently energy on the synchrotron process during their advection with the mixed pulsar-stellar wind. However, at larger energies the synchrotron energy losses should be taken into account. 

Relativistic electrons, after isotropization by the magnetic field in the mixed pulsar-stellar wind, should lose efficiently energy also on the IC scattering of the stellar radiation. We estimate the optical depth for the IC scattering process in the Thomson regime for electrons at the distance of $D_{\rm mix}$ on,
\begin{eqnarray}
\tau_{\rm IC/T} = n_{\star}\sigma_{\rm T}c(D_{\rm mix}/w_{\rm env})\approx 4.7T_4^3R_{11}^2/(\tau_{\rm d}v_8w_8),
\label{eq9}
\end{eqnarray}
\noindent
where the density of stellar photons is $n_\star = 2.75\times 10^9T_4^3(R_{11}/(\tau_{\rm d}v_8)^2$ cm$^{-3}$, and $D_{\rm mix}/w_{\rm env}$
is the characteristic time scale spend by electrons at the distance $D_{\rm mix}$ from the binary system.
Note that the scattering in the Thomson regime is valid only for electrons with energies
below $m_{\rm e}^2c^4/3k_{\rm B}T\sim  100/T_4$ GeV. However, in the case of LS~5039, the optical depth for the $\sim$TeV electrons
should be still close to unity, keeping in mind that the Klein-Nishina cross section drops inversely proportional with the electron energy. We conclude that electrons with TeV energies should efficiently lose energy on the IC scattering of stellar radiation after being isotropized in the pulsar-stellar wind nebula. However,  in the massive binary systems of the LS~5039 type, electrons, moving rectilinear in the pulsar wind in the outward direction from the companion star, are not expected to lose significantly energy already in the binary system due to the small scattering angles between electrons and stellar photons.
The $\gamma$-rays produced at the distance $D_{\rm mix}$ from the star
should escape from stellar radiation field without significant absorption
(e.g. see the scaling formula given by Eq.~26 in \cite{bed09} for the optical depths in the stellar radiation field).

The nebulae formed by the mixed pulsar-stellar winds can have very complicated structures in the case of binary systems in which the wind of the companion star is non-spherically-symmetric or the orbit has a large eccentricity. For example, in the case of the eccentricity
not far from unity the distance, $D_{\rm mix}$, differs significantly for the periastron and apastron passages. These distances can be estimated from, $D_{\rm mix}^{\rm per}\sim \tau_{\rm per}v_{\rm w}$, for the periastron passage and from $D_{\rm mix}^{\rm apo}\sim \tau_{\rm apo}v_{\rm w}$ for the apastron, where $\tau_{\rm per}$ and $\tau_{\rm apo}$ are the time scales describing the passage of the compact object 
through a half of the orbit centered on the periastron and apastron, respectively. In such eccentric binaries, nebulae should have the shape corresponding to the shape of the orbit of the companion stars.

In the present paper we outline only qualitatively the 
scenario describing the process and properties of the mixing of stellar and pulsar winds and concentrate on the expected high energy radiation produced in such scenario. The details of the mixing  process of the winds are very complicated. However, the first results of hydrodynamic numerical simulations of collisions of stellar winds are consistent with general features of the scenario described in this section (see e.g.~Pittard~2009, Lamberts et al.~2012).

\begin{table*}
  \caption{Basic parameters of stars and pulsars in binary systems: the radius of the companion star $R_\star$, its surface temperature $T_\star$, the mass loss rate, ${\dot M}$, and the velocity, $v_{\rm w}$, of the stellar wind, the magnetic field strengths at the surface of the star $B_\star$, the binary orbital period $\tau_{\rm orb}$ and the spin down luminosity of the observed (or supposed) pulsar $L_{\rm pul}$. }
  \begin{tabular}{lllllllllllll}
\hline 
\hline 
Name   &  $R_\star$ (cm) &  $T_\star$ (K) &  ${\dot M}$ $(M_\odot/\mathrm{yr})$ 
    &  $v_{\rm w}$ ($\mathrm{km/s}$) & $B_\star$ (G) (?) & $\tau_{\rm orb}$ (days) 
& $L_{\rm pul}$ (erg/s)  \\
\hline
B1259-63  & $4.2\times 10^{11}$  &   $2.7\times 10^4$  &  $2\times 10^{-7}$   & $2\times 10^3$  &  $10^2-10^3$ & 1236.7 & $8.3\times 10^{35}$\\
B1957+20   &  $10^{11}$  &  $8\times 10^3$  &  $3\times 10^{-10}$   &  $7\times 10^2$  & $1-10^3$  & 0.38  & $7.5\times 10^{34}$  \\
J1816+4510   &  $7\times 10^{10}$  &  $2\times 10^4$  &  $3\times 10^{-10}$   &  $7\times 10^2$  & $1-10^3$  & 0.36  & $5\times 10^{34}$  \\
LS~5039       &  $6.5\times 10^{11}$   &   $3.9\times 10^4$  &  $2.6\times 10^{-7}$ & $2.4\times 10^3$   & $10^2-10^3$ & 3.9  & $10^{37}$ \\
\hline
\hline 
\end{tabular}
  \label{tab1}
\end{table*}

%
%
\section{Nebulae around specific binary systems}

As an example, we estimate the parameters of nebulae around a few binary systems containing (or expected to contain) energetic pulsars, i.e.
classical radio pulsar PSR 1259-63, millisecond pulsar B1957+20 in the Black Widow type binary system, millisecond pulsar J1816+4510 in the Redback type binary system, and supposed classical pulsar in the binary system LS~5039 belonging  to the class of the TeV $\gamma$-ray binaries. The basic parameters of these binaries, collected or proposed in different works, are summarised in Table~1. 

Considered above binary systems are expected to be surrounded by nebulae with quite different parameters. We estimate the basic parameters of these nebulae  based on the formulae derived in Sect.~2 (see Table~2). The best conditions for $\gamma$-ray production seems to be provided by the binary system of the redback type, containing millisecond pulsar J1816+4510, and the supposed pulsar within the binary system LS~5039 from which already GeV and TeV $\gamma$-ray emission has been detected. The observed emission from LS~5039 is modulated with the binary period. Therefore, it is expected to be originated within the binary system. We propose that additional steady $\gamma$-ray emission component should be also produced in the nebula surrounding this system provided that it contains energetic pulsar. We calculate the steady $\gamma$-ray fluxes from the nebulae surrounding two binary systems mentioned above selecting them as the best candidates for production of observable steady $\gamma$-ray emission in the surrounding nebulae, i.e. those containing millisecond pulsar J1816+4510 and LS~5039.

\subsection{J1816+4510}

The millisecond pulsar J1816+4510, with the period of 3.2 ms, was recently discovered in the survey as a part of the Green Bank North Celestial Cap \citep{st12}. 
This pulsar is in the binary system with the low mass companion (mass below $0.16M_\odot$) having an orbital period of 8.7 hrs.
The distance to this binary has been estimated on 2.4 kpc, based on the
dispersion measure \citep{ka12}. The pulsed $\gamma$-ray emission from J1816+4510 has been also detected by {\it Fermi}, implying an energy conversion efficiency of $\sim 25\%$ from the pulsar's spin down luminosity into $\gamma$-rays. The pulsations have two peaks separated by $\sim 0.5$ in phase. The $\gamma$-ray spectrum is measured between 0.5 GeV and 5 GeV.
The spectrum is flat. It can be described by a pure power law with a photon index $-2.20\pm 0.07$ or a power law with an index of $-2.0\pm0.1$ and an exponential cut-off at $7.5\pm4.0\,$GeV \citep{ka12}. The companion star has a radius of $0.1R_\odot$ and the surface temperature in the range $1.8-2.1\times 10^4$ K \citep{ka12}. We have selected this binary as one of the interesting examples for our farther analysis due to the large spin down luminosity of the pulsar, $\sim 5\times 10^{34}$ erg s$^{-1}$ and a very hot companion, which is unusual for the Redback type millisecond binary system.

The likely parameters of this binary system allows us to estimate
the shape of the shock in this binary defined by the parameter $\eta$.
For $\eta\sim 1.2$ the shock almost divide the volume of the binary in two equal parts. For such geometry, the mixed wind dominates almost at the whole sphere around the binary system, i.e. $\Delta_{\rm mix}\sim 1$. 
The energies of electrons accelerated in this system can reach TeV energies
(see Table 2) for the intermediate magnetic field strengths from the range mentioned in Table 1. The acceleration process of these electrons will be considered in a more detail in the next section.
Also the optical depths for electrons at the distance of $D_{\rm mix}$ are in this case of the order of a few see Table~2). 

\begin{table*}
  \caption{Parameters of the nebulae around binary systems and energies of particles which produce $\gamma$-rays: the parameter $\eta$ defining the shock structure, the part of the hemisphere, $\Delta_{\rm mix}$, in which the mixed wind is confined,
the wind mixing distance $D_{\rm mix}$, the magnetic field $B_{\rm mix}$ at $D_{\rm mix}$, and the bulk speed of the mixed wind $w_{\rm env}$, the acceleration efficiency of electrons $\chi = 10^{-3}\chi_{-3}$, the optical depth for IC scattering of stellar photons $\tau_{\rm IC/T}$, and the electron acceleration limit due to advection, $E_{\rm adv}^{\rm max}$, and synchrotron $E_{\rm adv}^{\rm max}$ energy losses.}
\begin{tabular}{lllllllllll}
\hline 
\hline 
Name  & $\eta$  & $\Delta_{\rm mix}$ & $D_{\rm mix}$ (cm) & $B_{\rm mix}$ (G) &  $w_{\rm env}$ (cm/s) & $\chi_{-3}$ & $\tau_{\rm IC/T}$  & $E_{\rm syn}^{\rm max}$ (TeV) &  $E_{\rm adv}^{\rm max}$ (TeV) \\
\hline
B1259-63    & 0.01  &  0.09 & $2\times 10^{16}$  & $2\times 10^{-5} - 2\times 10^{-4}$  & $3.5\times 10^8$ & 0.14 & 0.4 &  47-150 &  1.7-17 \\
B1957+20  & 1.8   &  0.82 & $2.3\times 10^{12}$ & $4.4\times 10^{-4} - 0.4$ & $2.7\times 10^9$ & 8.1 & 0.4  & 17-460  &   $0.1-85$  \\
J1816+4510    & 1.2  & 0.95  &  $2\times 10^{12}$   & $4\times 10^{-4} - 0.4$ & $2.2\times 10^9$ & 5.4 &  3.4  &  12-410  &  0.05-50 \\
LS~5039    & 0.08 & 0.32 &  $8\times 10^{13}$    & $8\times 10^{-3} - 8\times 10^{-2}$ & $1.1\times 10^9$ & 1.3  & 110 &  7-24  &  7.7-77  \\
\hline
\hline 
\end{tabular}
  \label{tab2}
\end{table*}

%
\subsection{LS~5039}

LS~5039 is a well known binary system containing O type massive star and a compact object on the orbit with a period of 3.9 days
\citep{ca05}. The distance to the system is $2.9\pm 0.8$ kpc \citep{ca12}. The vicinity of LS~5039 shows radio morphology which indicates a presence of an energetic pulsar in this system \citep{mol12} as earlier suggested by \cite{dub06}. This binary system has been detected in the TeV $\gamma$-rays \citep{ah05} and also at GeV energies by {\it Fermi} \citep{ab09}, showing persistent emission modulated with the orbital period of the binary system. The GeV and TeV $\gamma$-ray light curves are clearly anticorrelated. The $\gamma$-ray spectrum extends up to $\sim 20$ TeV indicating the presence of multi-TeV particles within this binary system. The spectral slope changes dramatically with the orbital phase between -1.85 and -2.53 \citep{ah06}. We have selected this binary due to relatively small eccentricity ($e\sim 0.35$) and the well measured modulated TeV $\gamma$-ray emission which should allow to put constraints on the acceleration rate of relativistic electrons in terms of our model.

For this binary system, the shock structure bends around the pulsar, $\eta\sim 0.08$ (see Table 2). This means that the whole pulsar wind is overtaken by the stellar wind. The mixed pulsar-stellar wind
propagate only in a part of the sphere. Electrons are expected to be accelerated to several TeV for the 
likely surface magnetic field strength of the companion star.
The optical depth for electrons in the radiation field of the companion star (in the Thomson regime) is very large (of the order of $\sim$100) at the distance of $D_{\rm mix}$. Thus, efficient production of $\gamma$-rays is expected even from electrons with energies of the order of several TeV, which already scatter stellar radiation in the Klein-Nishina regime.

\section{Relativistic particles in nebulae}

We assume that relativistic particles in the nebula surrounding binary system are directly injected from the pulsar inner magnetosphere (close to mono-energetic case) or they are accelerated in the shocks created in pulsar-stellar wind collisions.
The Lorentz factors of the mono-energetic particles, $\gamma_{\rm e}^{\rm pul}$, are usually considered to be of the order of a few times $10^{\rm (4-6)}$.
It is expected that a significant part of these particles, those injected in the outward direction from the companion star, escape to the surrounding nebula and reach the distance of $D_{\rm mix}$. We estimate the maximum energies of particles which can be
accelerated at the turbulent region at the distance of $D_{\rm mix}$ by comparing their acceleration time scale and their advection time scale from the shock region given by Eq.~6. The acceleration time scale is given by, 
\begin{eqnarray}
\tau_{\rm acc} = E_{\rm e}/{\dot P}_{\rm acc}\approx 100E_{\rm TeV}/(\chi_{-3}B)~~~{\rm s},
\label{eq10}
\end{eqnarray}
\noindent
where ${\dot P}_{\rm acc} = \chi cE/R_{\rm L}$ and $R_{\rm L}$ is the Larmor radius of particles. The acceleration process is parametrised in this case by the acceleration efficiency $\chi = 10^{-3}\chi_{-3}$ which is assumed to be of the order of $\sim (w_{\rm env}/c)^2$.
Applying Eq.~4, we estimate the maximum energies of electrons due to their advection from the acceleration region on,
\begin{eqnarray}
E_{\rm adv}^{\rm max}\approx 10\chi_{-3}B_2R_{11}/w_8~~~{\rm TeV}.
\label{eq11}
\end{eqnarray}
\noindent
The maximum energies of particles, accelerated at the shock at the distance of $D_{\rm mix}$, can be additionally constrained by their energy losses. Therefore, we compare the acceleration time scale of particles with their energy loss time scale.
For the radiation energy loss time scales of electrons in the vicinity of the stars (synchrotron and IC processes), the standard formulae \citep{bg70} are applied. The comparison of the acceleration and the synchrotron energy loss time scales gives us the maximum energies of electrons due to the synchrotron energy losses,
\begin{eqnarray}
E_{\rm syn}^{\rm max}\approx 19(\chi_{-3}D_{13}/B_2R_{11})^{1/2}~~~{\rm TeV}.
\label{eq12}
\end{eqnarray}
\noindent
The maximum energies due to the Inverse Compton energy losses (in the Thomson regime) are given by,
\begin{eqnarray}
E_{\rm IC/T}^{\rm max}\approx 0.4(\chi_{-3}B_2D_{13}/R_{11})^{1/2}/T_4^2~~~{\rm TeV}.
\label{eq13}
\end{eqnarray}
\noindent
We stress that this second limit is only important in the Thomson regime, i.e. for electrons with energies $<100/T_4$ GeV.
For larger energies, due to the falling of the Klein-Nishina cross-section, the limit is relaxed.
The values of the maximum energies of electrons due to their escape with the expansion of the nebula and due to the synchrotron energy losses are shown for the considered binary systems in Table.~2. We conclude that electrons can be accelerated to energies estimated by the lower value between $E_{\rm syn}^{\rm max}$ and $E_{\rm adv}^{\rm max}$.
Note that these maximum energies are close to those values expected 
for electrons injected from the inner pulsar magnetosphere. Since
it is not clear which mechanism accelerate electron in the vicinity of the binary system, we consider two limiting cases: (1)  electrons are injected into the nebula surrounding the binary system with a power law spectrum extending between
$E_{\rm e}^{\rm min} = m_{\rm e}\gamma_{\rm e}^{\rm pul}$ and $E_{\rm e}^{\rm max}$, or (2) they are injected with monoenergetic energies $E_{\rm e}^{\rm min}$.

Relativistic hadrons are also expected to be accelerated in such nebulae. The maximum energies of these hadrons should be also limitted by their advection from the nebula. So then, they are given by the same formula as for electrons (Eq.~\ref{eq11}). 
We estimate the optical depth for the interaction of these relativistic hadrons with the matter of the mixed pulsar-stellar wind. It is assumed that hadrons are captured by the magnetic field
of the mixed wind above $D_{\rm mix}$. They are advected with the velocity of the wind $w_{\rm env}$. In such a case the optical depth is equal to
\begin{eqnarray}
\tau_{\rm pp} = \int_{D_{\rm mix}}^\infty n_{\rm H}\sigma_{\rm pp}{{c}\over{w_{\rm env}}}dD\approx 3\times 10^{-3}{{M_{-7}}\over{w_8^2D_{13}}},
\label{eq14}
\end{eqnarray}
\noindent
where $n_{\rm H} = {\dot M}/(4\pi D^2w_{\rm env}m_{\rm p})\approx 3\times 10^7M_{-7}/(D_{13}^2w_8)$ cm$^{-3}$ is the density of the matter in the mixed pulsar-stellar wind as a function of the distance from the star, $\sigma_{\rm pp} = 3\times 10^{-26}$ cm$^2$ is the cross section for proton-proton interaction, and $m_{\rm p}$ is the proton mass. For the parameters of the binary systems considered above (see Tables 1 and 2), the optical depths for hadronic interactions are clearly lower than unity. Therefore, we conclude that 
relativistic hadrons are not able to find enough target mass within the considered nebulae around binary systems to efficiently lose energy on hadronic processes. These protons mainly escape from the binary system and contribute to the galactic cosmic rays. Note that the stellar wind (and also mixed pulsar-stellar wind) can be quite inhomogeneous. However, relativistic protons have to be very efficiently captured within the regions with density effectively enhanced by a factor of $\sim$10$^{5-6}$, in respect to average densities estimated above for the considered binaries,
in order to provide their efficient interaction with the matter. 
Such efficient capturing seems to be rather unlikely in the stellar winds
since the filling factor of the stellar wind with such dense regions should be very low.

\section{Gamma-ray production in nebulae}

We calculate the $\gamma$-ray spectra produced by relativistic electrons within the nebulae in the Inverse Compton scattering of the stellar radiation and the Microwave Background Radiation. We assume that relativistic electrons are isotropized in the mixed pulsar-stellar wind at the distances above $D_{\rm mix}$. The maximum energies of the electrons are considered in the range which is consistent with the estimations reported in the last two columns in Table~2. The electrons are captured by the magnetic field of the mixed winds
and move gradually with the velocity of these winds. We also include the synchrotron energy losses of the electrons 
in the magnetic field of the winds with the values equal to $B_{\rm mix}$
at the distance $D_{\rm mix}$. 
This magnetic field drops with the distance, $D$, from the binary system as $\propto 1/D$. For example, in the case of J1816+4510, we apply the value
$B_{\rm mix} = 10^{-2}$ G. For this value, the energy density of stellar radiation at the distance $D_{\rm mix}$ overcomes the energy density of the magnetic field by a factor of $\sim$200. Therefore, the synchrotron energy losses can only start to be important at energies above a few hundred GeV, where the IC emission is suppressed by the Klein-Nishina cross-section provided that we consider the upper range of the magnetic fields (equal to $\sim 0.4\,$G, see Table~2). The relative importance of the synchrotron energy losses in respect to the IC energy losses, in the case of LS~5039 is similar to that
estimated for J1816+4510. The surface temperature of the companion star
in LS~5039 is a factor of two higher but the distance of the mixing region 
from the star (in units of stellar radius) is a factor of 4 larger.
Therefore the radiation field in the case of both sources at the distance 
$D_{\rm mix}$ are comparable. Note however, that the time scale for 
the IC scattering is a factor of $\sim$40 larger for LS~5039 which affects the optical depth for electrons (see $D_{\rm mix}$ and $\tau_{\rm IC/T}$ in Table 2).

\begin{figure*}
\includegraphics[width=0.95\textwidth]{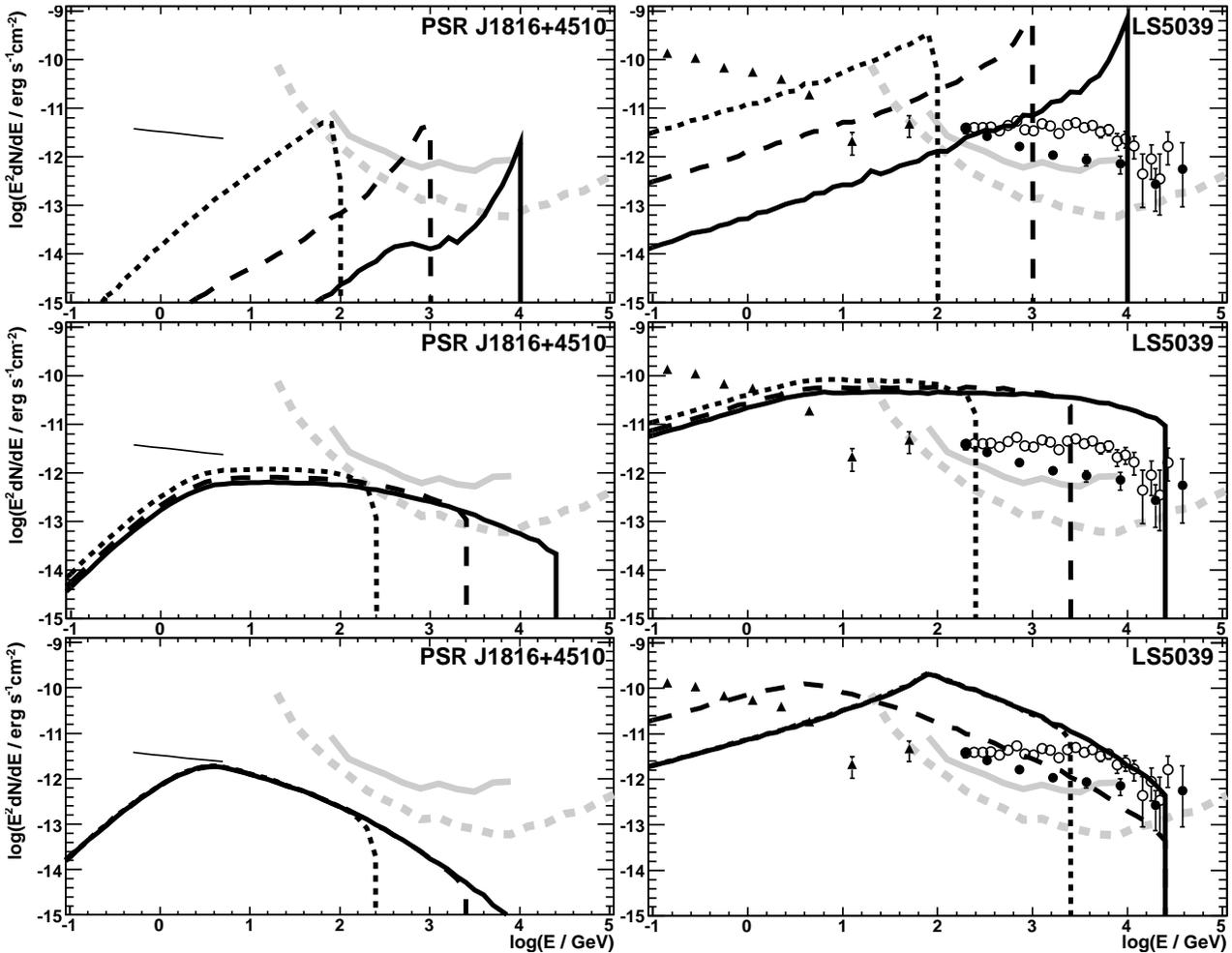}
\caption{
$\gamma$-ray SEDs expected from the nebula surrounding the Redback millisecond binary system J1816+4510 (on the left) and the binary system LS~5039 supposed to contain a classical pulsar (on the right). 
The upper panels show the results for monoenergetic injection of electrons with energy: $10^2$ GeV (dotted), $10^3$ GeV (dashed), and $10^4$ GeV (solid).
The middle panels show the results for a power-law injection with spectral index of $-2.1$ between $E_{\rm min} = 10$ GeV and $E_{\rm max} = 3\times 10^2$ GeV (dotted), $3\times 10^3$ GeV (dashed), and $3\times 10^4$ GeV (solid). 
The bottom panels show the results for a power-law injection with spectral index of $-3$ in energy ranges:
$10-3\times 10^2$ GeV (for J1816+4510) and $100-3\times 10^3$ GeV (for LS~5039) (dotted), 
$10-3\times 10^3$ GeV (for J1816+4510) and $10-3\times 10^4$ GeV (for LS~5039) (dashed), 
$10-3\times 10^4$ GeV (for J1816+4510) and $100-3\times 10^4$ GeV (for LS~5039) (solid).
It is assumed that $\Delta_{\rm mix}\cdot \epsilon = 10\%$ of the pulsar wind power is converted to relativistic electrons. 
In the case of J1816+4510 most of the relativistic electrons in the  pulsar wind escapes from the binary system but in the case of LS~5039 only a part of the electrons in the wind, estimated on $\Delta_{\rm IC}^{\rm bin}\sim 0.2$, can escape from the binary system.
For the comparison, the 100 hr differential sensitivity of MAGIC \citep{al12} and CTA \citep{ber12} are also marked by the gray solid and dotted curve respectively.
Full and empty data circles show the SED of LS~5039 measured by H.E.S.S. in 2 different orbital phases \citep{ah06} and the average spectrum from 30 months of the observations with \textit{Fermi}-LAT is shown with triangles \citep{had12}.
The thin black line in the left panels show the spectral fit to the \textit{Fermi}-LAT observations of J1816+4510 \citep{ka12}. 
}
\label{fig2}
\end{figure*}

We apply the Monte Carlo method in order to obtain the $\gamma$-ray spectra produced by electrons which IC scatter stellar photons. These electrons are advected from the binary system with the velocity of the mixed pulsar-stellar wind $w_{\rm env}$. The turbulent magnetic field in the mixed pulsar-stellar wind isotropizes the directions of electrons. Therefore, in the reference frame of electrons the stellar radiation from the companion star is isotropic. 
We take into account that electrons propagate in the stellar radiation field and the magnetic field which decrease with the distance from the binary system. The $\gamma$-ray spectra are calculated applying the formulae for the IC spectra in the general case from Blumenthal \& Gould (1970).

Those two binary systems, J1816+4510 and LS~5039, are selected for more detailed calculations. They seem to be the best candidates for the production of detectable $\gamma$-ray fluxes from the nebulae due to the luminous stellar companions and relatively compact nebulae. We assume that relativistic electrons take $\epsilon = 10\%$ of the rotational power lost by the pulsars.
In the case of the monoenergetic injection of electrons from the pulsar magnetosphere in the binary system LS~5039, only electrons, moving within the solid angle directed outward from the companion star, $\Delta\Omega_{\rm IC}^{\rm bin}$, can escape from the inner region of the binary system without significant 
energy losses on the IC process. 
These electrons escape in the outward direction (in respect to the star) since they move in the highly anisotropic stellar radiation field.
Based on the previous calculations of the radiation processes within the binary system LS~5039 \citep{bed06}, we estimate this part of the hemisphere on $\Delta_{\rm IC}^{\rm bin} = \Delta\Omega_{\rm IC}^{\rm bin}/4\pi \sim 0.2$. Thus, only $20\%$ of the relativistic electrons escape without significant energy losses to the outer regions of the binary system LS~5039. This additional factor has been included in our calculations for LS~5039. In the case of J1816+4510, electrons moving rectilinear through the binary system do not lose efficiently energy on the IC process due to lower surface temperature and smaller radius of the companion star. Therefore, we assume $\Delta_{\rm IC}^{\rm bin}\sim 1$ for J1816+4510.

The $\gamma$-ray spectra, expected from the nebulae around these two  binary systems, are shown in Fig.~2 for the case of the monoenergetic spectrum of electrons (upper figures) and the power law spectra of electrons which were ejected with the spectral indices equal to 2.1 (middle figures) and 3 (bottom figures). The parameters defining these spectra are reported in the figure caption. The maximum energies of electrons are consistent with the limits derived for
the case of dominant synchrotron energy losses or advection time scale with the mixed winds (Table~2).

\begin{table*}
  \caption{The upper limits on the product of the part of the hemisphere overtaken by the mixed pulsar-stellar wind ($\Delta_{\rm mix}$) and the energy conversion efficiency ($\epsilon$) from the pulsar to relativistic electrons for the binary system LS~5039 obtained from  comparison with H.E.S.S. and \textit{Fermi}-LAT spectral measurements}
  \begin{tabular}{l|lll|lll|lll}
\hline 
 & \multicolumn{3}{|l|}{Monoenergetic} & \multicolumn{3}{|l|}{Power-law, $\alpha=-2.6$} & \multicolumn{3}{|l|}{Power-law, $\alpha=-3$} \\\hline
$E_{\rm max}$ [GeV] & $10^2$ & $10^3$ & $10^4$  &  $10^{2.5}$ & $10^{3.5}$ & $10^{4.5}$  & $10^{3.5}$ & $10^{4.5}$ & $10^{4.5}$ \\
$E_{\rm min}$ [GeV] & $10^2$ & $10^3$ & $10^4$  &  $10$ & $10$ & $10$  & $100$  & $10$ & $100$\\ \hline\hline
$\Delta_{\rm mix}\cdot \epsilon$ &  $<0.3\%$ & $<0.05\%$ & $<0.03\%$  &  $<0.3\%$ & $<0.3\%$ & $<0.2\%$  & $<0.4\%$ & $<0.3\%$ & $<0.4\%$\\\hline
\end{tabular}
  \label{tab3}
\end{table*}

The $\gamma$-ray spectra, expected at the observer, are compared with the sensitivity of the present generation (e.g. MAGIC) and future (Cherenkov Telescope Array, CTA) imaging atmospheric Cherenkov telescopes (IACTs). 
Detection of the considered here TeV $\gamma$-ray emission from the Redback millisecond binary J1816+4510 with the present instruments is unlikely.
However, with a deep exposure of 100 hr, such emission could be detected by CTA as long as the energy conversion efficiency, $\epsilon$, is at least 10\%.
The $\gamma$-ray spectrum obtained from the \textit{Fermi}-LAT observations of that source does not constrain the product of part of the hemisphere subtracted by the mixed wind and the energy conversion efficiency for this source, i.e. $\Delta_{\rm mix}\cdot \epsilon$. 

The $\gamma$-ray spectra expected from the LS~5039 are $\sim$2 orders of magnitude above the expected sensitivity of CTA. 
With the assumed energy conversion efficiency from the pulsar to relativistic electrons ($10\%$), such emission should be clearly detected even by the presently operating  Cherenkov telescopes. 
The TeV emission observed from LS~5039 by the H.E.S.S. telescopes is modulated with the orbital period.
On the other hand, the emission expected from the presented here model should contribute as a constant component in the whole emission of the system.
Therefore, we can use the low state observed by H.E.S.S. telescopes to constrain the parameters which determine the energy transferred to the TeV electrons which escape from the binary system. This power depends on the product of a part of the hemisphere in which the mixed stellar-pulsar wind
is confined, $\Delta_{\rm mix}$ and the  energy conversion efficiency, $\epsilon$.
Derived upper limits on $\Delta_{\rm mix}\cdot \epsilon$ are reported in Table.~3 for different spectra of electrons in the case of LS~5039.
For the broad range of the model parameters, $\Delta_{\rm mix}\cdot \epsilon$ is severely constrained to be $\ll 1\%$.

It is believed that the energy conversion efficiency from the pulsar to relativistic electrons, in the inner pulsar magnetosphere and in the shock acceleration scenario, is of the order of $10\%$. Then, the part of the hemisphere in which the mixed wind is confined should be also below $\Delta_{\rm mix}\sim 10\%$. As shown in Sect.~2, this factor depends on the half opening angle of the shock within the binary system which is determined by the value of the parameter $\eta$. For the value of $\eta$, calculated for the supposed shock structure in LS~5039, $\Delta_{\rm mix}$ is larger than required above (see Table 2). 
In fact, different phenomena can effect the estimated value of $\Delta_{\rm mix}$.
At first, the value of $\eta$ might be much larger than $\sim$0.08, i.e the stellar wind may not effectively confine the pulsar wind. 
In fact the winds around Be type massive stars are expected to be aspherical with the dense and slow equatorial wind and the fast and rare poloidal wind.
If the pulsar is mainly immersed in the poloidal wind, than the shock structure might even bend around the star (see calculations in  \citealp{sb08}), i.e. the value of $\eta$ is larger than unity. 
In such case only a part of the pulsar wind can mix with the stellar wind.
Second, the electrons could be injected anisotropically from the pulsar or accelerated anisotropically in the pulsar wind. In fact, the pulsar winds are expected to be highly anisotropic with the dominant equatorial component (e.g. \citealp{bk02,vo08}). 
Such aspherical pulsar wind will introduce significant complications to the discussed scenario. More energy in relativistic electrons will be directed to the equatorial regions of the binary system resulting in much larger effective value of $\eta$ (even larger than unity). However the solid angle, in which mixed stellar-pulsar wind is confined, will become lower. Simple estimations, which base on the formulae in Sect.~2,
show that the increase of significant effective power of the wind in the equatorial direction results in more power transferred in the form of relativistic electrons to the nebula. 
However, in such a case a significant part of the particles, accelerated by the pulsar towards the companion star, could not be able to escape from the binary system due to efficient IC interactions with the stellar radiation. Therefore, the final effect
of the anisotropic pulsar wind on the energy transferred to nebula in the form of relativistic particles is difficult to estimate without 
more detailed calculations. 
The third possibility is that more electrons could be accelerated in the general direction towards the companion star. This might happen in the case of a significant collimation of the relativistic particles in the pulsar winds by the shock structure (e.g. see \citealp{du10,bo08}.
Then, electrons lose efficiently energy on IC process before they manage to escape into the mixed wind region. 

We conclude that the derived limits on the parameter $\Delta_{\rm mix}\cdot \epsilon$, determining the high energy emission from the nebula around LS~5039, indicate on either different parameters of the stellar wind than usually expected in the literature, or on the importance of anisotropic stellar and pulsar winds, or on the anisotropic injection/acceleration of relativistic electrons within the binary system.

\section{Conclusions}

We considered nebulae around binary systems which contain rotation powered pulsars and hot companion stars with a relatively large mass loss rate.
Such nebulae differ significantly from the pulsar wind nebulae around isolated rotation powered pulsars since a part of (or whole) the pulsar wind becomes loaded with the mass from the companion star. As a result, the pulsar wind relativistic plasma slows down and mixes with the matter of the stellar wind. The mixed pulsar-stellar wind moves from the binary system in the outward direction with a reduced velocity determined by the power of the pulsar wind and the mass of the stellar wind. 
We argue that the distance at which these two winds have to mix completely is determined by the orbital period of the binary system. 
This mixing process can concern the whole pulsar wind (for $\eta < 1$ the stellar wind pressure dominates over the pulsar wind pressure) or only a part of the pulsar wind for $\eta > 1$. In the latter case, the rest of the pulsar wind (along polar regions of the binary system) expands freely interacting with the interstellar medium at large distances from the binary system as observed in nebulae around isolated pulsars.

Due to the efficient mixing, relativistic electrons are isotropized in the reference frame of the mixed wind relatively close to the binary system. Therefore, they can efficiently comptonize thermal radiation from the companion star. This is in contrary to the nebulae around isolated pulsars where the inner part of the nebula is non-radiative due to the almost radial motion of the plasma from the pulsar. 

We calculate the $\gamma$-ray spectra produced by leptons in the IC scattering process in the case of the mono-energetic particles (leaving the light cylinder radius of the pulsar) and in the case of the power law spectra of leptons formed as a result of re-acceleration at the shocks produced in the pulsar-stellar wind collisions. Note that this $\gamma$-ray emission should be steady, i.e. independent on the phase of the binary system.   
As an example, we show the $\gamma$-ray spectra for the supposed pulsar in the binary system LS~5039 and for the recently discovered Redback type millisecond binary system J1816+4510, containing relatively hot companion star. It is found that if electrons take a part of the pulsar wind energy loss rate defined by the product of the part of the hemisphere, in which the mixed pulsar-stellar wind expands, and for the acceleration efficiency of electrons $\Delta\Omega\cdot \epsilon$ equal to $10\%$, than the TeV $\gamma$-ray emission from the millisecond binary system is comparable to the sensitivity of the Cherenkov Telescope Array (CTA). However, in the case of LS~5039, electrons with such value of $\Delta\Omega\cdot \epsilon$ should produce $\gamma$-ray emission on the level clearly above the sensitivity of both the present Cherenkov telescopes and CTA. By comparing the $\gamma$-ray spectra predicted for LS~5039 with the TeV $\gamma$-ray emission in the low state established by the H.E.S.S. telescopes, and with the average \textit{Fermi}-LAT observations, we put strong limits on $\Delta_{\rm mix}\cdot \epsilon <1\%$. Such low limit suggest that
either the parameters of the stellar wind are less extreme than considered up to now in the literature or the injection of electrons from the inner pulsar magnetosphere and/or accelerated at the shock is highly anisotropic. Most of relativistic electrons might be directed towards the companion star and lose efficiently energy already within the binary system.

It seems that the TeV $\gamma$-ray binary containing classical radio pulsar PSR B1259-63 should be also a good candidate source for the application of the above considered model. However, this pulsar is on a very elongated orbit around the companion star. Therefore, the nebula around this binary system is expected to have complicated, non-spherically symmetric shape. In this paper we provided calculations only for the binary systems which are expected to produce nebulae which can be more or less approximated by a symmetric shape. The more complicated case of the nebula around binary system, such as that containing PSR B1259-63, will be discussed in detail in a future paper.

\section*{Acknowledgments}
This work is supported by the grants from the Polish MNiSzW through the NCN No. 2011/01/B/ST9/00411 and through the NCBiR No. ERA-NET-ASPERA/01/10. 


\label{lastpage}
\end{document}